\documentclass{jps-cp}
\providecommand{\newblock}{\hskip .11em plus .33em minus .07em}

\title{Strangeness is the key: from $\bar{K}N$ to $\bar{D}_s D K$}

\author{Li-Sheng  \textsc{Geng}$^{1,2,3,4}$, Ming-Zhu \textsc{Liu}$^{5,6}$, and Jia-Ming \textsc{Xie}$^{1,7}$}

\inst{%
$^{1}$\quad School of Physics, Beihang University, Beijing 102206, China\\
$^{2}$\quad Sino-French Carbon Neutrality Research Center, \'Ecole Centrale de P\'ekin/School of General Engineering, Beihang University, Beijing 100191, China\\

$^{3}$\quad Peng Huanwu Collaborative Center for Research and Education, Beihang University, Beijing 100191, China\\
$^{4}$\quad  Southern Center for Nuclear-Science Theory (SCNT), Institute of Modern Physics, Chinese Academy of Sciences, Huizhou 516000, China\\
$^{5}$ \quad
Frontiers Science Center for Rare Isotopes, Lanzhou University,
Lanzhou 730000, China\\
$^{6}$\quad  School of Nuclear Science and Technology, Lanzhou University, Lanzhou 730000, China\\
$^{7}$ \quad Department of Physics, Graduate School of Science, The University of Tokyo, Tokyo 113-0033, Japan
}

\email{lisheng.geng@buaa.edu.cn}

\recdate{February 19, 2026}

\abst{ The kaon, the lightest hadron containing a strangeness quark, is very peculiar. It is a Nambu-Goldstone boson, but significantly heavier than the pion. As a result, its interaction with a matter particle, such as the nucleon or a heavy-light meson, such as the $D$ meson, is completely determined by chiral dynamics and much stronger than its pion cousin.  The strong attractive interaction has brought us many surprises and is manifested in the peculiar nature of many particles, such as the mysterious $\Lambda(1405)$ and $D_{s0}^*(2317)$. These two particles can be understood as $\bar{K}N$ and $DK$ hadronic molecules, respectively. They also imply the existence of three-body hadronic molecules that await future discovery. In this talk, I review some recent developments in our understanding of hadronic interactions involving the kaon.  }

\kword{Kaon interaction, $\Lambda(1405)$, $D_{s0}^*(2317)$, three-body hadronic molecules, genuine three-body forces \ldots}

\begin{document}
\maketitle
\section{Motivation: Exotic Hadrons as Hadronic Molecules}
One central question in physics addresses the fundamental building blocks of nature across scales and the rules governing their interactions. For instance, nucleons are bound together to form nuclei by the residual strong interaction--the nuclear force. Nuclei attract electrons via electromagnetic interactions and form atoms. Through theresidual electromagnetic interaction, atoms attract one another, thereby giving rise to various molecular structures. They are the building blocks of our visible universe. As of today, the smallest strongly interacting matter particles are quarks. According to the standard model of particle physics, they interact via gluons to form color-singlet hadrons. The constituent quark model (CQM), proposed by Gell-Mann and Zweig in 1964~\cite{Gell-Mann:1964ewy,Zweig:1964jf,Zweig:1964ruk}, classifies hadrons into two categories: mesons as pairs of quark and antiquark ($q\bar{q}$), such as the pion, the kaon, and the $D$ meson, and baryons of three quarks ($qqq$), such as the proton and the neutron, collectively referred to as nucleons. 

Beginning in 2003, such a simple but mysteriously successful picture was challenged by the discovery of the so-called \textbf{exotic hadrons}, i.e.,  hadrons that do not fit easily into the CQM $q\bar{q}$ or $qqq$ states. They can be classified into compact tetraquark states, compact pentaquark states, hadronic molecules, glueballs, hybrids, or even kinematic cusps.  An interesting observation is that many exotic hadrons, if not all, lie near two-body hadronic thresholds, suggesting a hadronic molecular nature, such as the familiar deuteron, a weakly bound state of a neutron and a proton with a binding energy of about 2.2 MeV.

\begin{figure}[ttt]
\centering
\includegraphics[scale=0.45]{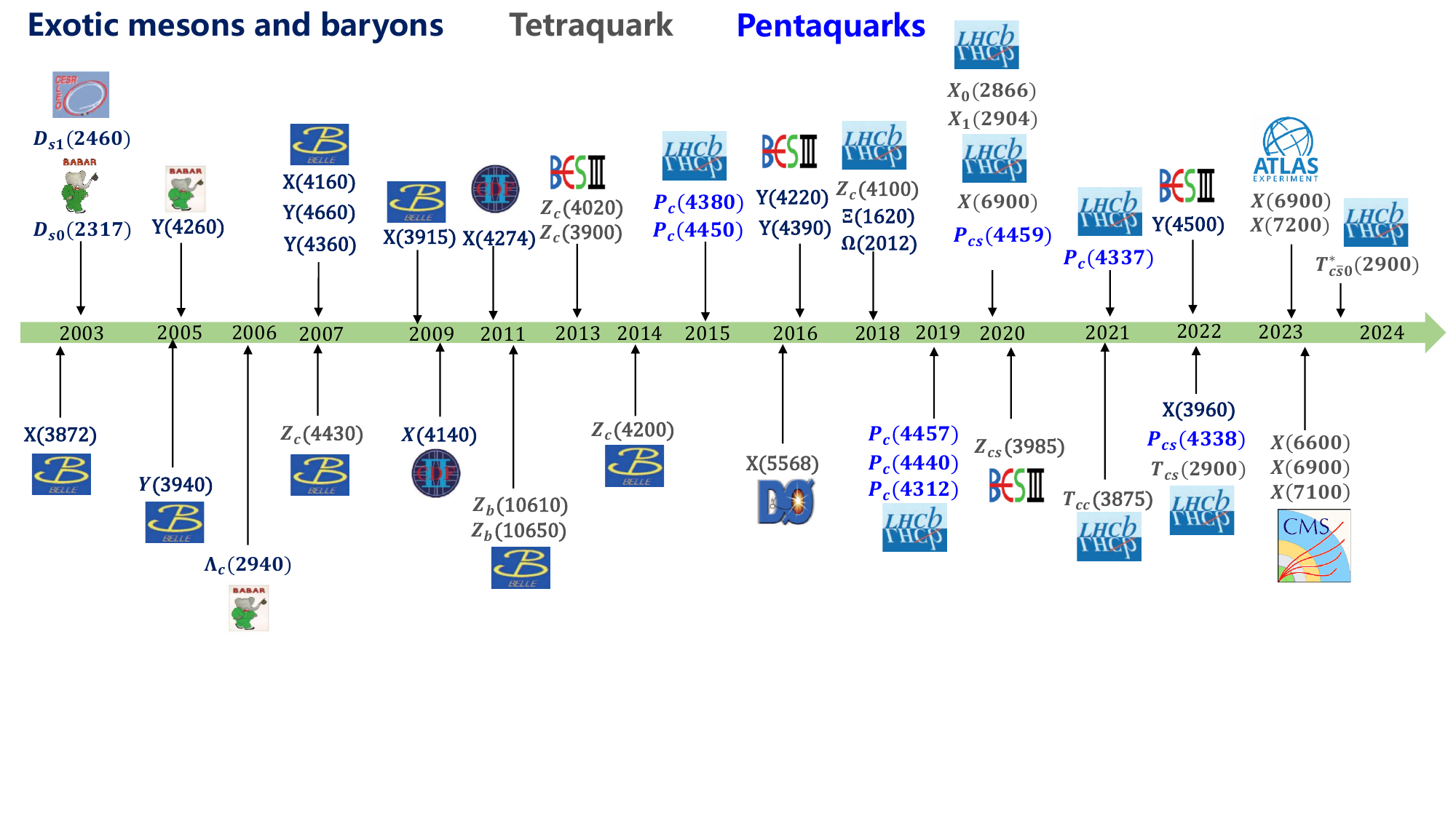}
 \vspace{-1.8cm}
\caption{Discovery timeline of selected exotic hadrons. The color code denotes their classification: ``exotic baryons and mesons'' denote that they have quantum numbers of CQM hadrons, but have peculiar properties; ``tetraquarks'' denote mesons that have a minimum of four constituent quarks/antiquarks; ``pentaquarks'' denote baryons that contain at least five constituent quarks.}
\label{f1}
\end{figure}
Fig. 1 shows a selection of these hadronic states and their discovery time.  We will not discuss each of them and refer interested readers to the many excellent reviews~\cite{Chen:2016qju,Lebed:2016hpi,Oset:2016lyh,Esposito:2016noz,Dong:2017gaw,Guo:2017jvc,Ali:2017jda,Karliner:2017qhf,Guo:2019twa,Liu:2024uxn,Wang:2025sic,Doring:2025sgb}. 

\section{$\Lambda(1405)$ as a $\bar{K}N$ Bound State and Its Two-Pole Structure}
The $\Lambda(1405)$, discovered in 1961~\cite{Alston:1961zzd}, can be considered the first exotic hadronic state because it is significantly lighter than its nucleon counterpart, $N^*(1535)$, though it contains a heavier strangeness quark. The $\Lambda(1405)$ was predicted to be a $\bar{K}N$ bound state even before its experimental discovery~\cite{Dalitz:1959dn}. Such a picture received strong support in the chiral unitary approaches that combine SU(3)$_L$$\times$SU(3)$_R$  chiral dynamics and elastic unitarity (see Refs.~\cite{Oller:2000ma,Oller:2019opk,Mai:2020ltx} for a more complete list of references).  An unexpected finding of the chiral unitary approaches is that the $\Lambda(1405)$ actually corresponds to two dynamically generated poles~\cite{Oller:2000fj,Jido:2003cb}, between the $\pi \Sigma$ and $\bar{K}N$ thresholds. Such a two-pole picture was recently confirmed in the unified description of meson-baryon scattering at next-to-next-to-leading order~(NNLO)~\cite{Lu:2022hwm} and by lattice QCD simulations~\cite{BaryonScatteringBaSc:2023zvt,BaryonScatteringBaSc:2023ori}. In Ref.~\cite{Xie:2023cej}, we showed how the two-pole structure emerges from the underlying chiral dynamics, namely, the universal Weinberg-Tomozawa interaction, the large mass difference between the kaon and the pion, and the SU(3) flavor symmetry breaking of the $\bar{K}N$ and $\pi\Sigma$ channels. We also argued that given the universality of the Weinberg-Tomozawa interaction, such a two-pole structure is expected in other systems as well, such as the $K_1(1270)$~\cite{Roca:2005nm,Geng:2006yb} and the $\Xi(1890)$~\cite{Molina:2023uko}. 

In Fig.2, we show the trajectories of the two poles of the $\Lambda(1405)$. The evolution of the higher pole is simple. As the pion mass increases, both its real and imaginary parts decrease. This indicates that the effective $\bar{K}N$ interaction and coupling to $\pi\Sigma$ decrease as the pion (kaon) mass increases. Note that as the pion mass increases, the two thresholds also increase. On the other hand, the trajectory of the lower pole is more complicated and highly nontrivial. As the pion mass increases, it first becomes a virtual state from a resonant state for a pion mass of about 200 MeV. For a pion mass of about 300 MeV, it becomes a bound state and remains so up to the pion mass of 500 MeV. The evolution of the lower pole clearly demonstrates the chiral dynamics underlying the two-pole structure of the $\Lambda(1405)$. Such a highly nontrivial quark mass dependence of the two poles can be deemed as key to deciphering the nature of the two-pole structure. In a recent work, we similarly studied the trajectories of the two poles of the $K_1(1270)$~\cite{Xie:2025xew}.
Following the lattice QCD studies~\cite{BaryonScatteringBaSc:2023zvt,BaryonScatteringBaSc:2023ori}, several works have examined the quark-mass dependence of the $\Lambda(1405)$~\cite{Zhuang:2024udv,Ren:2024frr,Guo:2023wes}. These studies yield largely consistent conclusions, but a question remains. That is, whether, at NLO, the higher and lower poles exchange their flavor-structure assignments~\cite{Zhuang:2024udv,Guo:2023wes}. In Ref.~\cite{Liu:2024tvt}, it was argued that one can use the hadronic
decays of charmonia into $\bar{\Lambda}\Sigma\pi$ and $\bar{\Lambda}\Sigma\pi$ as flavor filters to single out the flavor octet and singlet poles. It will be interesting to check what happens at the next-to-next-to-leading order~\cite{Lu:2022hwm}. 
We note that many UChPT studies have predicted the existence of the isovector counterparts of the $\Lambda(1405)$~\cite{Oller:2000fj,Ikeda:2012au,Guo:2012vv,Cieply:2016jby,Khemchandani:2018amu,Lu:2022hwm,Pittler:2025upn}, which await future experimental confirmation. 

\begin{figure}[htpb]
\centering
\includegraphics[scale=0.3]{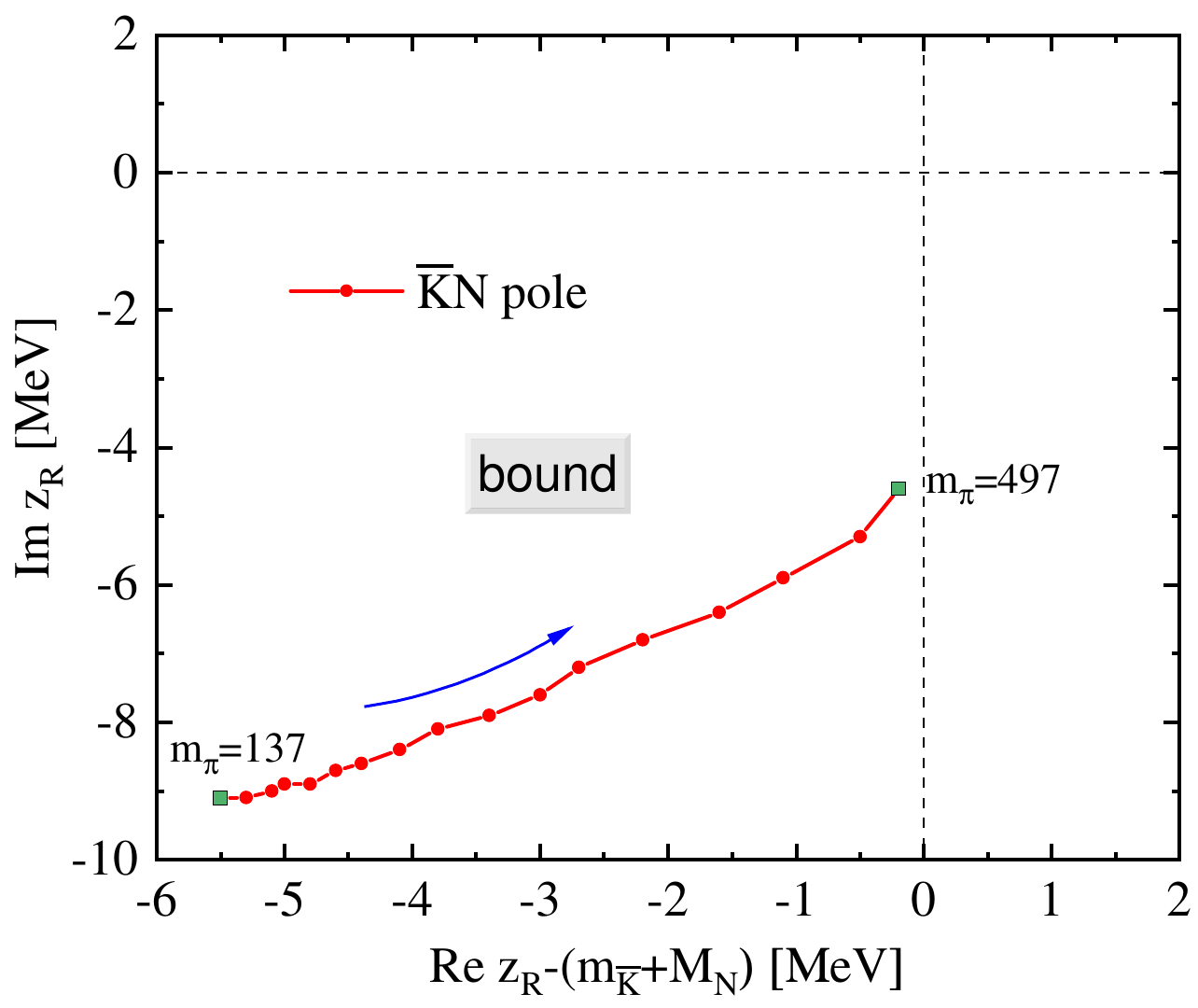}\quad\quad
\includegraphics[scale=0.3]{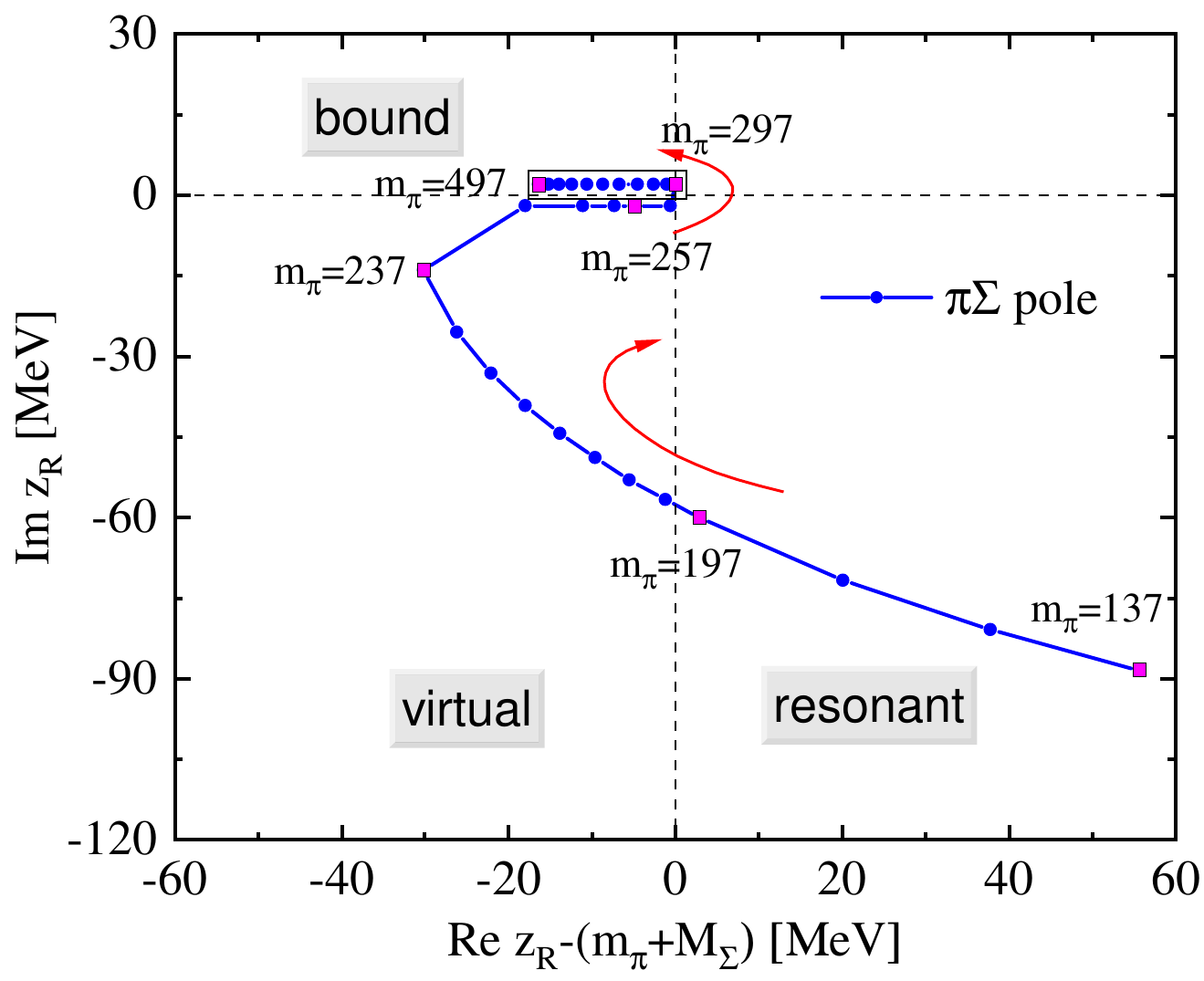}
\caption{Trajectories of the  two poles of $\Lambda(1405)$ as functions of the pion mass  $m_{\pi}$ from 137 MeV to 497 MeV. Critical masses are labeled by solid squares, with points equally spaced between them. Taken from Ref.~\cite{Xie:2023cej}. }
\label{poletrejec}
\end{figure}

\section{$D_{s0}^{*}(2317)$ as a $DK$ Molecule and the Unique $\bar{D}_{s}DK$ Three-Body State}

 $D_{s0}^*(2317)$ was first discovered by the BaBar Collaboration~\cite{Aubert:2003fg} and then confirmed by the CLEO~\cite{Besson:2003cp} and Belle~\cite{Krokovny:2003zq} collaborations. It is located  45 MeV below the $DK$ threshold and has a decay width less than 3.8 MeV. The observed mass and width are far from the predicted mass of 2460 MeV and the width of hundreds of MeV in the quark model~\cite{Barnes:2003dj}. Thus, $D_{s0}^*(2317)$ is difficult to be interpreted as a conventional $c\bar{s}$ state. On the other hand, due to the strongly attractive $DK$ interaction, it can  be easily explained as a $DK$ molecule~\cite{Barnes:2003dj,Kolomeitsev:2003ac,Hofmann:2003je,Guo:2006fu,Gamermann:2006nm,Liu:2012zya,Altenbuchinger:2013vwa,Mohler:2013rwa,Lang:2014yfa,Bali:2017pdv,Yang:2013rwa}. 
 
 It is interesting to note that in the unitarized chiral approach, the $D^*K$ interaction is the same as the $DK$ interaction up to heavy quark spin symmetry breaking effects. As a result, the existence of a $DK$ molecule implies the existence of a $D^*K$ molecule. In Ref.~\cite{Altenbuchinger:2013vwa}, fixing the next-to-leading order low-energy constants (LECs) and a subtraction constant by fitting to the lattice QCD  scattering lengths~\cite{Liu:2012zya}, and then solving the Bethe-Salpeter equation, one found two poles in the strangeness $-1$, which coincide with the experimentally known $D_{s0}^*(2317)$ and $D_{s1}(2460)$. In such a picture, one can easily understand the fact that the mass difference between $D_{s1}(2460)$ and $D_{s0}^*(2317)$ is almost the same as the mass difference between $D^*$ and $D$, because in the molecule picture, the mass difference originates from the different masses of the constituents, as the interactions between the constituents are the same due to heavy quark spin symmetry.  
 
If the $D_{s0}^*(2317)$ is dominantly a $DK$ bound state, it is natural to ask what happens if one adds one $D/\bar{D}/\bar{D}^*$ into the $DK$ pair. Will the resulting three-body systems bind? If they do, what are the binding energies and strong decay widths? Where can future experiments search for them?  Many such systems have been studied previously, such as the $DDK$~\cite{Wu:2019vsy}, $D\bar{D}K$~\cite{Wu:2020job,Wei:2022jgc}, $D\bar{D}^*K$~\cite{Wu:2020job,Ma:2017ery,Ren:2018pcd}, $D^*DK$~\cite{Tan:2024omp}, and $DDKK$~\cite{Pan:2023zkl} systems.  See Refs.~\cite{Wu:2022ftm,Liu:2024uxn} for reviews.

\begin{figure}[ttt]
  \centering
 \includegraphics[width=0.8\textwidth]{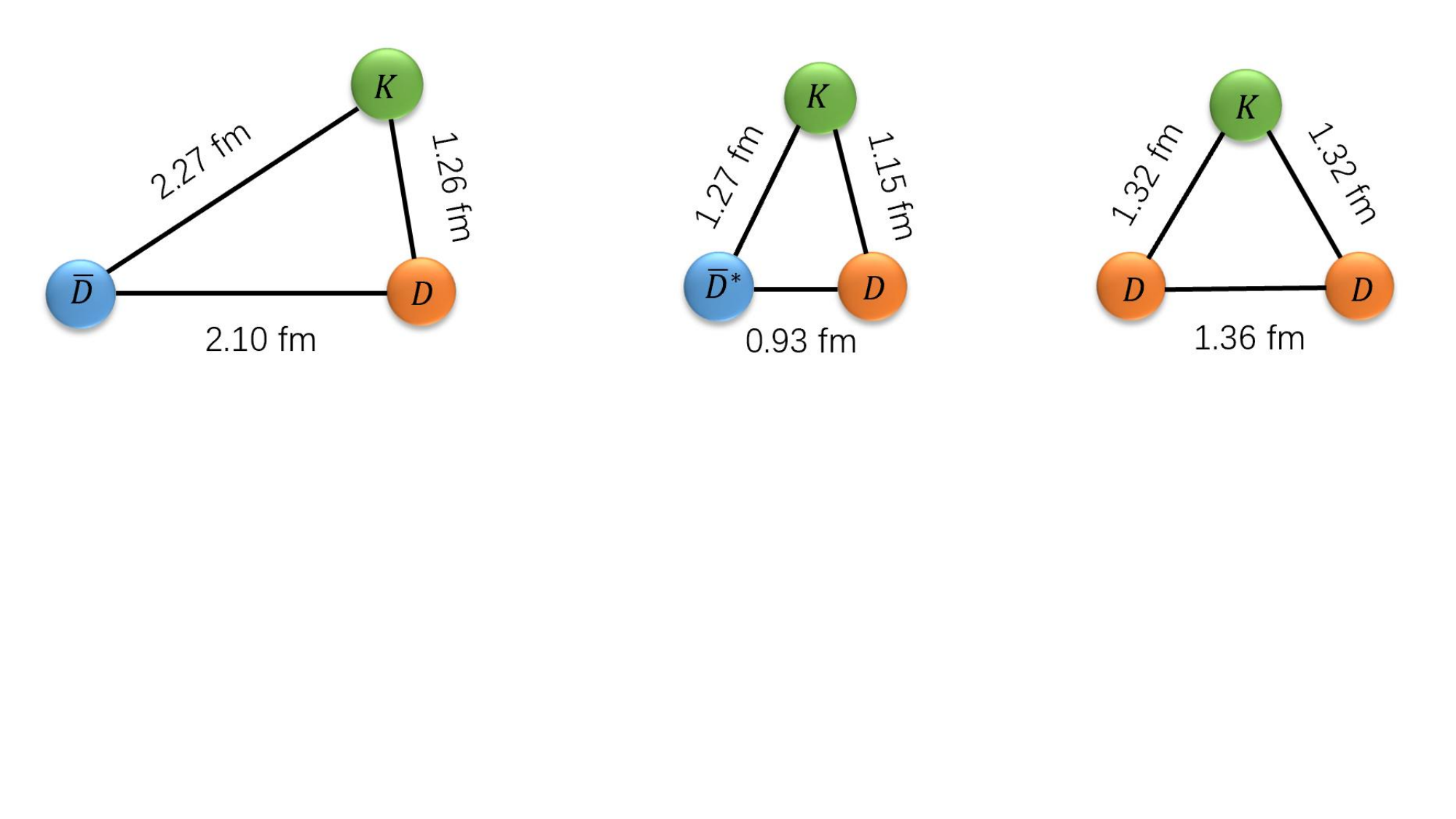}
 \vspace{-3.6cm}
  \caption{ RMS radii of subsystems in the $D\bar{D}K$ (left), $D\bar{D}^*K$ (middle), and $DDK$~\cite{Wu:2019vsy}  bound states with a cutoff $R_c=1.0$ fm. Taken from Ref.~\cite{Wu:2020job}.}
  \label{RMS}
\end{figure}

The binding energies of the $DDK$, $D\bar{D}K$, and $D\bar{D}^*K$ bound states are $67.1\sim71.2$, $48.9^{+1.4}_{-2.4}$, and $77.3^{+3.1}_{-6.6}$~MeV, respectively. Among the three bound states, the $D\bar{D}^*K$ state has the largest binding energy, because the $D\bar{D}^*$ interaction is attractive enough to generate $X(3872)$ dynamically. The uncertainties arise from the consideration of the likely existence of a short-range repulsion and from the use of several cutoffs (see Refs.~\cite{Wu:2019vsy,Wu:2020job} for more details). In Fig.~\ref{RMS}, we show the root mean square (RMS) radii of the two-body subsystems of the three bound states.  Consistent with the binding energies, the $D\bar{D}K$ system is more extended, while the $D\bar{D}^*K$ state is more compact. One should note that the spatial distributions are more sensitive to the details of the two-body potentials, and the results shown in Fig.~\ref{RMS} are for illustrative purposes only.

The Belle collaboration has searched for the $DDK$ state in $e^+e^-$ collisions, but has obtained only upper limits on the product of branching fractions~\cite {Belle:2020xca}. In contrast to the open-charm $DDK$ state, the hidden-charm $D\bar{D}K$ state is more likely to be observed in $e^+e^-$ collisions. 
The possibility of detecting the $D\bar{D}K$ state in inclusive $e^+e^-\rightarrow c\bar{c}$ collisions has been studied in Ref.~\cite{Wu:2022wgn}.  As indicated in Ref.~\cite{Liu:2024uxn}, these three-body molecules can be produced in the exclusive $b$-flavored meson decays, i.e., $DDK$ in $B_c$ meson decays and $D\bar{D}K$ in $B$ meson decays.  

\subsection{Unique Features of the $\bar{D}_{s}DK$ Three-Body System}
\begin{figure}[!h]
	\centering
	\includegraphics[width=8.0cm]{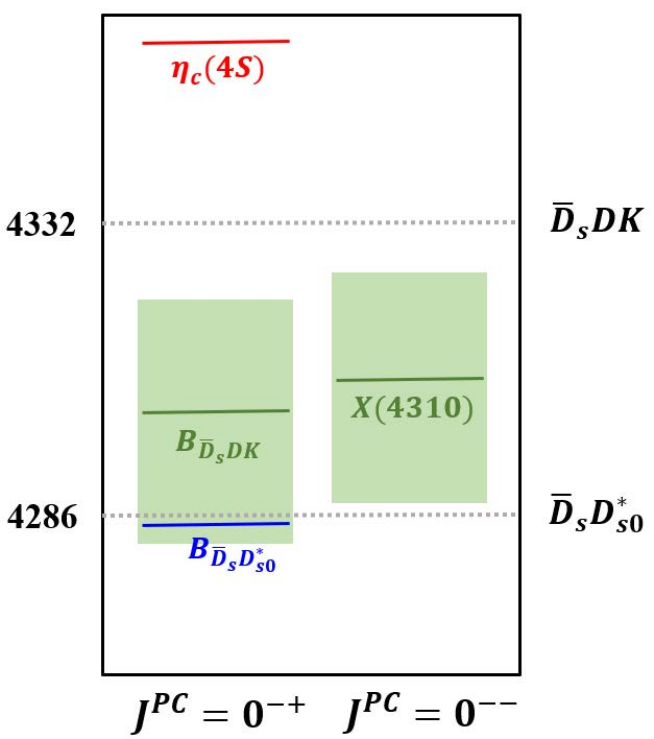}\\
	\caption{The predicted masses of $0^{--}$ $\bar{D}_s DK$ three-body molecular state and $0^{-+}$ $\bar{D}_s DK$ three-body molecular state compared with the $0^{-+}$ $\bar{D}_sD_{s0}^*$ molecular state and their thresholds. Taken from Ref.~\cite{Wu:2025fzx}.
               }\label{Spectrumdia}
\end{figure}

The $\bar{D}_{s}DK$ system is unique because of the following reasons. First,
there exists a $C$-parity-dependent interaction. Second, it has exotic quantum numbers, $J^{PC}=0^{--}$, implying no mixing with conventional $q\bar{q}$ or $qqq$ states. Third, there are no bound two-body subsystems, indicating a genuine three-body bound state, as shown in Fig.~\ref{Spectrumdia}.

The three-body wave function for the $\bar{D}_sDK$ system of good $C$ parity reads
\begin{equation}\label{schd}
\Psi^{C}=\frac{1}{\sqrt{2}}(\Psi_{\bar{D}_sDK}+C\Psi_{D_s\bar{D}\bar{K}}'),
\end{equation}
where  $C=\pm1$ represents the $C$-parity eigenvalue,  and $\Psi$ ($\Psi'$) denotes the wave function of the $\bar{D}_{s}DK$ (${D}_{s}\bar{D}\bar{K}$) system.
The wave function $\Psi^C$ can be obtained by solving the Schr\"odinger equation with the Hamiltonian $H=T+T'+V+V'+V^C$,  where $T$ ($T'$) and $V$ ($V'$) are the kinetic energy term and the hadron-hadron potentials of $\Psi$ ($\Psi'$), respectively.
Since the strong interaction is invariant under charge conjugation, the Schr\"odinger equation of $\Psi^C$ can be simplified as
\begin{equation}\label{schd1}
\langle \Psi|(T+V_{DK}+V_{\bar{D}_sK}+V_{\bar{D}_sD}+V^C-E)|\Psi\rangle=0,
\end{equation}
which can be solved using the Gaussian Expansion Method (GEM)~\cite{Hiyama:2003cu}.

Here we employ the contact-range effective field theory (EFT) to construct the $DK$, $\bar{D}_sK$, and $D\bar{D}_s$ potentials~\cite{Hidalgo-Duque:2012rqv,Liu:2023cwk}. 
A Gaussian-shaped potential in coordinate space can parameterize the contact-range potential in momentum space 
\begin{equation}
\label{TBpoten}
    V(r)=C
    _a\frac{e^{(-r/R_c)2}}{\pi^{3/2}R_c^3},
\end{equation}
 where $R_c$ is a cut-off radius of the order of a typical hadronic size. The parameter $C_a$ characterizes the strength of two-body potentials, which are determined by fitting the masses of molecular candidates as well as symmetries~\cite{Guo:2006fu,Guo:2009ct,Liu:2012zya,Altenbuchinger:2013vwa,Ji:2022vdj}. 
 We assume the   $D_{s0}^*(2317)$  as a mixture of a $DK-D_s\eta$ molecular state and a $c\bar{s}$ bare state rather than a pure $DK$ molecule. 
The uncertainty in the extracted $DK$ potential is estimated by varying the molecular compositeness of the $D_{s0}^*(2317)$ from $50\%$ to $100\%$.
Finally, one can obtain the following approximate ratio for the contact potentials for a cutoff $\Lambda=1$~GeV: $C_a^{DK}:C_a^{\bar{D}_s K}: C_a^{\bar{D}_s D}=1:0.5:0.1$.

The potential $V^{C}$, dependent on the $C$-parity, is a three-body interaction correlating the wave functions $\Psi$ and $\Psi'$. Here, since the $DK$ potential can form a bound state ${D}_{s0}^*$, we use an effective two-body $\bar{D}_sD_{s0}^*$ potential utilizing the one-boson-exchange (OBE) model~\cite{Liu:2010hf,Shen:2010ky}.   The $\eta$ exchange potential for the $\bar{D}_sD_{s0}^*$ system is  
\begin{small}
\begin{eqnarray}
V^{C=\pm}&=& \mp\frac{2}{3}\frac{k^2}{f_\pi^2}q_0^2(\frac{e^{-mr}-e^{-\Lambda r}}{4\pi r}-\frac{\Lambda^2-m^2}{8 \pi \Lambda}e^{-\Lambda r}),
\end{eqnarray}
\end{small}
where $q_0=m_{D_{s0}^*}-m_{D_s}$, $k=0.56$, and $f_\pi=130$~MeV~\cite{Liu:2010hf}.

\begin{table}[ttt]
 \centering
 \caption{  Binding energy (in units of MeV), weights of Jacobi channels, root mean square radii (in units of fm), and expectation values of the Hamiltonian (in units of MeV) of the $0^{--}$ $\bar{D}_sDK$ molecule. Taken from Ref.~\cite{Wu:2025fzx}
 \label{threebody0--} }
 \begin{tabular}{c|  ccc c}
 \hline\hline
 Scenarios&  B.E.($0^{--})$ &    $ P_{\bar{D}_sK-D}$& $P_{DK-\bar{D}_s}$ & $ P_{\bar{D}_sD-K}$\\
 \hline
$\alpha=1$ & $22^{+23}_{-14}$& $11^{+1}_{-1}$ \%&$78^{-1}_{+2}$ \%&$11^{+0}_{-1}$ \%\\
 $\alpha=2$&  $20^{+22}_{-13}$& $10^{+1}_{-1} \%$& $80^{-1}_{+2}$ \%&$10^{+0}_{-1}$ \%\\
 \hline
  Scenarios&  $r_{\bar{D}_sK}$ &  $r_{DK}$  & $r_{\bar{D}_sD}$ &$\langle T \rangle$\\
 \hline
$\alpha=1$ & $1.6_{+0.9}^{-0.4}$& $1.2_{+0.5}^{-0.3}$  &$1.4_{+0.8}^{-0.3}$&$177_{-74}^{+81} $  \\
 $\alpha=2$& $1.7_{+1.4}^{-0.4}$& $1.2_{+0.7}^{-0.3}$   &$1.6_{+1.3}^{-0.5}$&$169_{-78}^{+82}$ \\
 \hline
  Scenarios& $ \langle V_{D_s\bar{K}}\rangle$~~~ &  $\langle V_{DK}\rangle~~~$  &   $ \langle V_{D_s\bar{D}}\rangle$&$ \langle V^{C=-}_{D_s\bar{D}^*_{s0}}\rangle$\\
 \hline
$\alpha=1$ & $-40_{+20}^{-25}$& $-147_{+62}^{-73}$&$-14_{+6}^{-7}$&$2_{-1}^{+0}$\\
 $\alpha=2$&  $-37_{+22}^{-25}$& $-143_{+64}^{-73}$&$-13_{+7}^{-6}$&$3_{-1}^{+2}$\\
 \hline
 \hline
 \end{tabular}
 \end{table}

In Table~\ref{threebody0--}, the $0^{--}$ $\bar{D}_sDK$ system is predicted to be a bound state with a binding energy of $21^{+24}_{-14}$ MeV.  Our results indicate that when the $DK-D_s\eta$ molecular component  of the $D_{s0}^*(2317)$ ranges from  $50\%$ to  $100\%$, the $\bar{D}_sDK$ system always remains bound. Such a three-body molecule is denoted by $X(4310)$.  The weight of each Jacobi channel $\Phi_i$  is calculated by $P_i=\langle\Psi|\hat{P}_i^G|\Psi\rangle$, with $|\Psi\rangle=\sum_i |\Phi_i\rangle$ and the generalized projection operator
    $\hat{P}_i^G=\sum_j |\Phi_i\rangle S^{-1}_{ij}\langle\Phi_j|$ suited for a non-orthogonal basis~\cite{PhysRevB.90.075128}, 
where $S_{ij}^{-1}$ is the element of the inverse of the overlap $S$-matrix of Jacobi channels with $S_{ij}=\langle\Phi_i|\Phi_j\rangle$ ($i,j=1-3$).
Interestingly, the weights of Jacobi channels $i=1-3$ in Fig.~\ref{RMS} are stable, which are about 10\%, 80\%, and 10\% of $\bar{D}_sK-D$, $DK-\bar{D}_s$, and $\bar{D}_sD-K$,  respectively, varying  only $1\sim 2$ percent. 
This indicates the  $0^{--}$ three-body bound state is dominated by the $(DK)-\bar{D}_s$ channel, weakly dependent on the molecular component of the $D_{s0}^*(2317)$. 
  The root mean square radii and expectation values of the Hamiltonian of the $0^{--}$ $\bar{D}_sDK$ bound state are also presented in Table~\ref{threebody0--}, thereby providing a more intuitive demonstration of the spatial extent of the predicted molecular state and the relative contributions of individual interaction terms in the Hamiltonian.

 To facilitate potential observation of the predicted state $X(4310)$, we have investigated its decay properties and production mechanisms through triangle diagrams, focusing particularly on $B$ decays~\cite{Cheng:2004ru,Faessler:2007gv}.  We assume that the $B$ meson firstly decays into a  $\bar{D}^*$ meson and a $D_{s0}^*$ hadronic molecule, and then the $\bar{D}^*$ meson scatters into a $\bar{D}_s$ and a $K$ meson. Finally, the $\bar{D}_sDK$ molecule is dynamically generated by the subsystem $\bar{D}_sD_{s0}^*$. 
Our results indicate that the $X(4310)$ dominantly decays into $\bar{D}^*D$.  
Using a similar approach, we estimated the branching fraction of the decay $B \to K X(4310)$ is approximately $10^{-6}$.  In the isospin limit,  we estimate the branching fractions of the decays $B^+ \to [X(4310) \to {D}^{*\pm}D^\mp ] K^+$ to be  
$5\times 10^{-7}$. Referring to the  branching fractions  $\mathcal{B}(B^+ \to {D}^{*\pm}D^\mp  K^+ )=6\times 10^{-4}$~\cite{BaBar:2010tqo,LHCb:2020qdy}, we estimate the   ratios of $\mathcal{B}[B^+ \to (X(4310) \to {D}^{*\pm}D^\mp ) K^+]/\mathcal{B}(B^+ \to {D}^{*\pm}D^\mp  K^+) \sim 10^{-3}$. The event number of the  decays $B^+ \to {D}^{*\pm}D^\mp  K^+$ of the LHCb Collaboration corresponding to an integrated luminosity of $9$~fb$^{-1}$ is around $2\times 10^{3}$~\cite{LHCb:2024vfz}. One can expect that the event number of the 
 decays $B^+ \to [X(4310) \to {D}^{*\pm}D^\mp ] K^+$  would reach  at least $10$ and  $10^2$  corresponding to  the integrated luminosity of $50$~fb$^{-1}$ and $350$~fb$^{-1}$. Therefore, we strongly recommend experimental searches for the  $0^{--}$ $\bar{D}_sDK$ molecule in the decay channels   $B^+ \to  {D}^{*\pm}D^\mp  K^+ $.

\section{Summary and Outlook}
We have provided a concise review on $\Lambda(1405)$, $D_{s0}^*(2317)$, and the pertinent $\bar{K}N$ and $DK$ interactions. Various studies show that both $\Lambda(1405)$ and $D_{s0}^{*}(2317)$ can be largely understood as hadronic molecules of $\bar{K}N$ and $DK$.  They can be dynamically generated via chiral dynamics. We argued that $\Lambda(1405)$'s two-pole structure is a universal feature of chiral dynamics, confirmed by the N$^2$LO calculations, lattice QCD, and experimental data. Based on the molecular picture of the $D_{s0}^*(2317)$, we predicted the existence of a unique $0^{--}\bar{D}_{s}DK$ three-body bound state $X(4310)$ and studied its production rates in B decays. Interestingly, the charge-parity-dependent interaction implies that such systems provide a unique platform for understanding genuine three-body interactions, which have remained elusive in nuclear physics for decades. In a recent study~\cite{Pan:2025vyg}, we showed that three-body forces play a more important role in determining whether the $1^{-+}$ $\bar{D}^*D\eta$ system is bound.

Looking ahead, we hope that some of the predictions, particularly the existence of three-body hadronic molecules, can inspire more theoretical and experimental studies. 

\section{Acknowledgments}
This work is partly supported by the National Key R\&D Program of China under Grant No. 2023YFA1606703 and the National Natural Science Foundation of China under Grant No. W2543006 and No. 12435007. M.Z.L acknowledges support from the National Natural Science Foundation of China under Grant No.~12575086.


\begin{thebibliography}{10}

\bibitem{Gell-Mann:1964ewy}
Murray Gell-Mann.
\newblock {A Schematic Model of Baryons and Mesons}.
\newblock {\em Phys. Lett.}, 8:214--215, 1964.

\bibitem{Zweig:1964jf}
G.~Zweig.
\newblock {\em {An SU(3) model for strong interaction symmetry and its
  breaking. Version 2}}, pages 22--101.
\newblock 2 1964.

\bibitem{Zweig:1964ruk}
G.~Zweig.
\newblock {An SU(3) model for strong interaction symmetry and its breaking.
  Version 1}.
\newblock {\em CERN-TH-401}, 1 1964.

\bibitem{Chen:2016qju}
Hua-Xing Chen, Wei Chen, Xiang Liu, and Shi-Lin Zhu.
\newblock {The hidden-charm pentaquark and tetraquark states}.
\newblock {\em Phys. Rept.}, 639:1--121, 2016.

\bibitem{Lebed:2016hpi}
Richard~F. Lebed, Ryan~E. Mitchell, and Eric~S. Swanson.
\newblock {Heavy-Quark QCD Exotica}.
\newblock {\em Prog. Part. Nucl. Phys.}, 93:143--194, 2017.

\bibitem{Oset:2016lyh}
Eulogio Oset et~al.
\newblock {Weak decays of heavy hadrons into dynamically generated resonances}.
\newblock {\em Int. J. Mod. Phys. E}, 25:1630001, 2016.

\bibitem{Esposito:2016noz}
A.~Esposito, A.~Pilloni, and A.~D. Polosa.
\newblock {Multiquark Resonances}.
\newblock {\em Phys. Rept.}, 668:1--97, 2017.

\bibitem{Dong:2017gaw}
Yubing Dong, Amand Faessler, and Valery~E. Lyubovitskij.
\newblock {Description of heavy exotic resonances as molecular states using
  phenomenological Lagrangians}.
\newblock {\em Prog. Part. Nucl. Phys.}, 94:282--310, 2017.

\bibitem{Guo:2017jvc}
Feng-Kun Guo, Christoph Hanhart, Ulf-G. Mei\ss{}ner, Qian Wang, Qiang Zhao, and
  Bing-Song Zou.
\newblock {Hadronic molecules}.
\newblock {\em Rev. Mod. Phys.}, 90(1):015004, 2018.

\bibitem{Ali:2017jda}
Ahmed Ali, Jens~S\"oren Lange, and Sheldon Stone.
\newblock {Exotics: Heavy Pentaquarks and Tetraquarks}.
\newblock {\em Prog. Part. Nucl. Phys.}, 97:123--198, 2017.

\bibitem{Karliner:2017qhf}
Marek Karliner, Jonathan~L. Rosner, and Tomasz Skwarnicki.
\newblock {Multiquark States}.
\newblock {\em Ann. Rev. Nucl. Part. Sci.}, 68:17--44, 2018.

\bibitem{Guo:2019twa}
Feng-Kun Guo, Xiao-Hai Liu, and Shuntaro Sakai.
\newblock {Threshold cusps and triangle singularities in hadronic reactions}.
\newblock {\em Prog. Part. Nucl. Phys.}, 112:103757, 2020.

\bibitem{Liu:2024uxn}
Ming-Zhu Liu, Ya-Wen Pan, Zhi-Wei Liu, Tian-Wei Wu, Jun-Xu Lu, and Li-Sheng
  Geng.
\newblock {Three ways to decipher the nature of exotic hadrons: Multiplets,
  three-body hadronic molecules, and correlation functions}.
\newblock {\em Phys. Rept.}, 1108:1--108, 2025.

\bibitem{Wang:2025sic}
Zhi-Gang Wang.
\newblock {Review of the QCD sum rules for exotic states}.
\newblock {\em Front. Phys. (Beijing)}, 21(1):016300, 2026.

\bibitem{Doring:2025sgb}
Michael D\"oring, Johann Haidenbauer, Maxim Mai, and Toru Sato.
\newblock {Dynamical coupled-channel models for hadron dynamics}.
\newblock {\em Prog. Part. Nucl. Phys.}, 146:104213, 2026.

\bibitem{Alston:1961zzd}
Margaret~H. Alston et~al.
\newblock {Study of Resonances of the $\Sigma$-$\pi$ System}.
\newblock {\em Phys. Rev. Lett.}, 6:698--702, 1961.

\bibitem{Dalitz:1959dn}
R.~H. Dalitz and S.~F. Tuan.
\newblock {A possible resonant state in pion-hyperon scattering}.
\newblock {\em Phys. Rev. Lett.}, 2:425--428, 1959.

\bibitem{Mai:2020ltx}
Maxim Mai.
\newblock {Review of the $\Lambda(1405)$: A curious case of a strangeness resonance}.
\newblock {\em Eur. Phys. J. ST}, 230(6):1593--1607, 2021.

\bibitem{Oller:2000ma}
J.~A. Oller, E.~Oset, and A.~Ramos.
\newblock {Chiral unitary approach to meson meson and meson - baryon
  interactions and nuclear applications}.
\newblock {\em Prog. Part. Nucl. Phys.}, 45:157--242, 2000.

\bibitem{Oller:2019opk}
J.~A. Oller.
\newblock {Coupled-channel approach in hadron--hadron scattering}.
\newblock {\em Prog. Part. Nucl. Phys.}, 110:103728, 2020.

\bibitem{Oller:2000fj}
J.A. Oller and Ulf~G. Meissner.
\newblock {Chiral dynamics in the presence of bound states: Kaon nucleon
  interactions revisited}.
\newblock {\em Phys. Lett. B}, 500:263--272, 2001.

\bibitem{Jido:2003cb}
D.~Jido, J.~A. Oller, E.~Oset, A.~Ramos, and U.~G. Meissner.
\newblock {Chiral dynamics of the two $\Lambda(1405)$ states}.
\newblock {\em Nucl. Phys. A}, 725:181--200, 2003.

\bibitem{Lu:2022hwm}
Jun-Xu Lu, Li-Sheng Geng, Michael Doering, and Maxim Mai.
\newblock {Cross-Channel Constraints on Resonant Antikaon-Nucleon Scattering}.
\newblock {\em Phys. Rev. Lett.}, 130(7):071902, 2023.

\bibitem{BaryonScatteringBaSc:2023zvt}
John Bulava et~al.
\newblock {Two-Pole Nature of the $\Lambda(1405)$ resonance from Lattice QCD}.
\newblock {\em Phys. Rev. Lett.}, 132(5):051901, 2024.

\bibitem{BaryonScatteringBaSc:2023ori}
John Bulava et~al.
\newblock {Lattice QCD study of $\pi\Sigma$-$\bar{K}N$ scattering and the
  $\Lambda(1405)$ resonance}.
\newblock {\em Phys. Rev. D}, 109(1):014511, 2024.

\bibitem{Xie:2023cej}
Jia-Ming Xie, Jun-Xu Lu, Li-Sheng Geng, and Bing-Song Zou.
\newblock {Two-pole structures as a universal phenomenon dictated by
  coupled-channel chiral dynamics}.
\newblock {\em Phys. Rev. D}, 108(11):L111502, 2023.

\bibitem{Roca:2005nm}
L.~Roca, E.~Oset, and J.~Singh.
\newblock {Low lying axial-vector mesons as dynamically generated resonances}.
\newblock {\em Phys. Rev. D}, 72:014002, 2005.

\bibitem{Geng:2006yb}
L.~S. Geng, E.~Oset, L.~Roca, and J.~A. Oller.
\newblock {Clues for the existence of two $K_1(1270)$ resonances}.
\newblock {\em Phys. Rev. D}, 75:014017, 2007.

\bibitem{Molina:2023uko}
R.~Molina, Wei-Hong Liang, Chu-Wen Xiao, Zhi-Feng Sun, and E.~Oset.
\newblock {Two states for the $\Xi(1820)$ resonance}.
\newblock {\em Phys. Lett. B}, 856:138872, 2024.

\bibitem{Xie:2025xew}
Jia-Ming Xie, Zhi-Wei Liu, Jun-Xu Lu, Haozhao Liang, Raquel Molina, and Li-Sheng Geng.
\newblock {Chiral Evolution and Femtoscopic Signatures of the $K_1(1270)$ Resonance}.
\newblock arXiv:2511.14380 [hep-ph], 2025.

\bibitem{Zhuang:2024udv}
Zejian Zhuang, Raquel Molina, Jun-Xu Lu, and Li-Sheng Geng.
\newblock {Pole trajectories of the $\Lambda(1405)$ help establish its dynamical nature}.
\newblock {\em Sci. Bull.}, 70:1953--1961, 2025.

\bibitem{Ren:2024frr}
Xiu-Lei Ren.
\newblock {Light-quark mass dependence of the $\Lambda(1405)$ resonance}.
\newblock {\em Phys. Lett. B}, 855:138802, 2024.

\bibitem{Guo:2023wes}
Feng-Kun Guo, Yuki Kamiya, Maxim Mai, and Ulf-G. Mei\ss{}ner.
\newblock {New insights into the nature of the $\Lambda(1380)$ and
  $\Lambda(1405)$ resonances away from the SU(3) limit}.
\newblock {\em Phys. Lett. B}, 846:138264, 2023.

\bibitem{Liu:2024tvt}
Xiao-Hai Liu, Ying-Bo He, Li-Sheng Geng, Feng-Kun Guo, and Ju-Jun Xie.
\newblock {Identifying the two-pole structure of $\Lambda(1405)$}.
\newblock {\em Phys. Rev. D}, 113:L051501, 2026.

\bibitem{Ikeda:2012au}
Yoichi Ikeda, Tetsuo Hyodo, and Wolfram Weise.
\newblock {Chiral SU(3) theory of antikaon-nucleon interactions with improved threshold constraints}.
\newblock {\em Nucl. Phys. A}, 881:98--114, 2012.

\bibitem{Guo:2012vv}
Zhi-Hui Guo and J.~A. Oller.
\newblock {Meson-baryon reactions with strangeness $-1$ within a chiral framework}.
\newblock {\em Phys. Rev. C}, 87(3):035202, 2013.

\bibitem{Cieply:2016jby}
A.~Ciepl\'y, M.~Mai, Ulf-G.~Mei{\ss}ner, and J.~Smejkal.
\newblock {On the pole content of coupled channels chiral approaches used for the $\bar{K}N$ system}.
\newblock {\em Nucl. Phys. A}, 954:17--40, 2016.


\bibitem{Khemchandani:2018amu}
K.~P. Khemchandani, A.~Mart{\'\i}nez Torres, and J.~A. Oller.
\newblock {Hyperon resonances coupled to pseudoscalar- and vector-baryon channels}.
\newblock {\em Phys. Rev. C}, 100(1):015208, 2019.

\bibitem{Pittler:2025upn}
Ferenc Pittler, Maxim Mai, Ulf-G.~Mei{\ss}ner, Ryan F. Ferguson, Peter Hurck, David G. Ireland, and Bryan McKinnon.
\newblock {Universal parameters of the $\Lambda(1380)$, the $\Lambda(1405)$, and their isospin partners from a combined analysis of lattice QCD and experimental results}.
\newblock {\em Phys. Rev. D}, 112(7):074037, 2025.


\bibitem{Aubert:2003fg}
B.~Aubert et~al.
\newblock {Observation of a narrow meson decaying to $D_s^+ \pi^0$ at a mass of
  2.32-GeV/$c^2$}.
\newblock {\em Phys. Rev. Lett.}, 90:242001, 2003.

\bibitem{Besson:2003cp}
D.~Besson et~al.
\newblock {Observation of a narrow resonance of mass 2.46-GeV/$c^2$ decaying to
  $D^{*+}_s \pi^0$ and confirmation of the $D^*_{sJ}(2317)$ state}.
\newblock {\em Phys. Rev. D}, 68:032002, 2003.
\newblock [Erratum: Phys. Rev.D75,119908(2007)].

\bibitem{Krokovny:2003zq}
P.~Krokovny et~al.
\newblock {Observation of the $D_{sJ}(2317)$ and $D_{sJ}(2457)$ in $B$ decays}.
\newblock {\em Phys. Rev. Lett.}, 91:262002, 2003.

\bibitem{Barnes:2003dj}
T.~Barnes, F.~E. Close, and H.~J. Lipkin.
\newblock {Implications of a $DK$ molecule at 2.32-GeV}.
\newblock {\em Phys. Rev. D}, 68:054006, 2003.

\bibitem{Kolomeitsev:2003ac}
E.~E. Kolomeitsev and M.~F.~M. Lutz.
\newblock {On Heavy light meson resonances and chiral symmetry}.
\newblock {\em Phys. Lett. B}, 582:39--48, 2004.

\bibitem{Hofmann:2003je}
J.~Hofmann and M.~F.~M. Lutz.
\newblock {Open charm meson resonances with negative strangeness}.
\newblock {\em Nucl. Phys. A}, 733:142--152, 2004.

\bibitem{Guo:2006fu}
Feng-Kun Guo, Peng-Nian Shen, Huan-Ching Chiang, Rong-Gang Ping, and Bing-Song
  Zou.
\newblock {Dynamically generated $0^+$ heavy mesons in a heavy chiral unitary
  approach}.
\newblock {\em Phys. Lett. B}, 641:278--285, 2006.

\bibitem{Gamermann:2006nm}
D.~Gamermann, E.~Oset, D.~Strottman, and M.~J. Vicente~Vacas.
\newblock {Dynamically generated open and hidden charm meson systems}.
\newblock {\em Phys. Rev. D}, 76:074016, 2007.

\bibitem{Liu:2012zya}
Liuming Liu, Kostas Orginos, Feng-Kun Guo, Christoph Hanhart, and Ulf-G.
  Meissner.
\newblock {Interactions of charmed mesons with light pseudoscalar mesons from
  lattice QCD and implications on the nature of the $D_{s0}^*(2317)$}.
\newblock {\em Phys. Rev. D}, 87(1):014508, 2013.

\bibitem{Altenbuchinger:2013vwa}
M.~Altenbuchinger, L.~S. Geng, and W.~Weise.
\newblock {Scattering lengths of Nambu-Goldstone bosons off $D$ mesons and
  dynamically generated heavy-light mesons}.
\newblock {\em Phys. Rev. D}, 89(1):014026, 2014.

\bibitem{Mohler:2013rwa}
Daniel Mohler, C.~B. Lang, Luka Leskovec, Sasa Prelovsek, and R.~M. Woloshyn.
\newblock {$D_{s0}^*(2317)$ Meson and $D$-Meson-Kaon Scattering from Lattice
  QCD}.
\newblock {\em Phys. Rev. Lett.}, 111(22):222001, 2013.

\bibitem{Lang:2014yfa}
C.~B. Lang, Luka Leskovec, Daniel Mohler, Sasa Prelovsek, and R.~M. Woloshyn.
\newblock {$D_s$ mesons with $DK$ and $D^*K$ scattering near threshold}.
\newblock {\em Phys. Rev. D}, 90(3):034510, 2014.

\bibitem{Bali:2017pdv}
Gunnar~S. Bali, Sara Collins, Antonio Cox, and Andreas Sch\"afer.
\newblock {Masses and decay constants of the $D_{s0}^*(2317)$ and
  $D_{s1}(2460)$ from $N_f=2$ lattice QCD close to the physical point}.
\newblock {\em Phys. Rev. D}, 96(7):074501, 2017.

\bibitem{Yang:2013rwa}
Zhi Yang, Guang-Juan Wang, Jia-Jun Wu, Makoto Oka, and Shi-Lin Zhu.
\newblock {Novel Coupled Channel Framework Connecting the Quark Model and Lattice QCD for the Near-threshold $D_s$ States}.
\newblock {\em Phys. Rev. Lett.}, 128(11):112001, 2022.

\bibitem{Wu:2019vsy}
Tian-Wei Wu, Ming-Zhu Liu, Li-Sheng Geng, Emiko Hiyama, and Manuel~Pavon
  Valderrama.
\newblock {$DK$, $DDK$, and $DDDK$ molecules\textendash{}understanding the
  nature of the $D_{s0}^*(2317)$}.
\newblock {\em Phys. Rev. D}, 100(3):034029, 2019.

\bibitem{Wu:2020job}
Tian-Wei Wu, Ming-Zhu Liu, and Li-Sheng Geng.
\newblock {Excited $K$ meson, $K_c(4180)$, with hidden charm as a $D\bar{D}K$
  bound state}.
\newblock {\em Phys. Rev. D}, 103(3):L031501, 2021.

\bibitem{Wei:2022jgc}
Xiang Wei, Qing-Hua Shen, and Ju-Jun Xie.
\newblock {Faddeev fixed-center approximation to the $D\bar{D}K$ system and the
  hidden charm $K_{c\bar{c}}(4180)$ state}.
\newblock {\em Eur. Phys. J. C}, 82(8):718, 2022.

\bibitem{Ma:2017ery}
Li~Ma, Qian Wang, and Ulf-G. Mei\ss{}ner.
\newblock {Double heavy tri-hadron bound state via delocalized $\pi$ bond}.
\newblock {\em Chin. Phys. C}, 43(1):014102, 2019.

\bibitem{Ren:2018pcd}
Xiu-Lei Ren, Brenda~B. Malabarba, Li-Sheng Geng, K.~P. Khemchandani, and
  A.~Mart\'\i{}nez~Torres.
\newblock {$K^*$ mesons with hidden charm arising from $KX(3872)$ and
  $KZ_c(3900)$ dynamics}.
\newblock {\em Phys. Lett. B}, 785:112--117, 2018.

\bibitem{Tan:2024omp}
Yue Tan, Xuejie Liu, Xiaoyun Chen, Youchang Yang, Hongxia Huang, and Jialun
  Ping.
\newblock {Dynamical study of $D^*DK$ and $D^*D\bar{D}$ systems at quark
  level}.
\newblock {\em Phys. Rev. D}, 110(1):016005, 2024.

\bibitem{Pan:2023zkl}
Ya-Wen Pan, Ming-Zhu Liu, Jun-Xu Lu, and Li-Sheng Geng.
\newblock {Systematic studies of $DDKK$ and
  $D\bar{D}K\bar{K}$ four-hadron molecules}.
\newblock {\em Phys. Rev. D}, 109(5):054026, 2024.

\bibitem{Wu:2022ftm}
Tian-Wei Wu, Ya-Wen Pan, Ming-Zhu Liu, and Li-Sheng Geng.
\newblock {Multi-hadron molecules: status and prospect}.
\newblock {\em Sci. Bull.}, 67:1735--1738, 2022.

\bibitem{Belle:2020xca}
Y.~Li et~al.
\newblock {Search for a doubly-charged $DDK$ bound state in $\Upsilon(1S,2S)$
  inclusive decays and via direct production in $e^+e^-$ collisions at
  $\sqrt{s}$ = 10.520, 10.580, and 10.867 GeV}.
\newblock {\em Phys. Rev. D}, 102(11):112001, 2020.

\bibitem{Wu:2022wgn}
Tian-Chen Wu and Li-Sheng Geng.
\newblock {Theoretical investigation of the molecular nature of $D_{s0}^*(2317)$ and
  $D_{s1}(2460)$ and the possibility of observing the $D\bar{D}K$ bound
  state $K_{c\bar{c}}(4180)$ in inclusive
  $e^+e^-\rightarrow c\bar{c}$ collisions}.
\newblock {\em Phys. Rev. D}, 108(1):014015, 2023.

\bibitem{Wu:2025fzx}
Tian-Wei Wu, Ming-Zhu Liu, and Li-Sheng Geng.
\newblock {Implication of the Existence of $J^{PC}=0^{--}$ $\bar{D}_sDK$ Bound State on the Nature of $D_{s0}^*(2317)$, and a New Configuration of Exotic State}.
\newblock {\em Phys. Rev. Lett.}, 135(3):031902, 2025.

\bibitem{Hiyama:2003cu}
E.~Hiyama, Y.~Kino, and M.~Kamimura.
\newblock {Gaussian expansion method for few-body systems}.
\newblock {\em Prog. Part. Nucl. Phys.}, 51:223--307, 2003.

\bibitem{Hidalgo-Duque:2012rqv}
C.~Hidalgo-Duque, J.~Nieves, and M.~Pavon Valderrama.
\newblock {Light flavor and heavy quark spin symmetry in heavy meson
  molecules}.
\newblock {\em Phys. Rev. D}, 87(7):076006, 2013.

\bibitem{Liu:2023cwk}
Ming-Zhu Liu, Xi-Zhe Ling, and Li-Sheng Geng.
\newblock {Productions of $D_{s0}^*(2317)$ and $D_{s1}(2460)$ in $B_{(s)}$ and
  $\Lambda_b$($\Xi_b$) decays}.
\newblock {\em Phys. Rev. D}, 109(5):056014, 2024.

\bibitem{Guo:2009ct}
Feng-Kun Guo, Christoph Hanhart, and Ulf-G. Meissner.
\newblock {Interactions between heavy mesons and Goldstone bosons from chiral
  dynamics}.
\newblock {\em Eur. Phys. J. A}, 40:171--179, 2009.

\bibitem{Ji:2022vdj}
Teng Ji, Xiang-Kun Dong, Miguel Albaladejo, Meng-Lin Du, Feng-Kun Guo, Juan
  Nieves, and Bing-Song Zou.
\newblock {Understanding the $0^{++}$ and $2^{++}$ charmonium(-like) states
  near 3.9 GeV}.
\newblock {\em Sci. Bull.}, 68:688--697, 2022.

\bibitem{Liu:2010hf}
Xiang Liu, Zhi-Gang Luo, and Shi-Lin Zhu.
\newblock {Novel charmonium-like structures in the $J/\psi\phi$ and
  $J/\psi\omega$ invariant mass spectra}.
\newblock {\em Phys. Lett. B}, 699:341--344, 2011.
\newblock [Erratum: Phys.Lett.B 707, 577 (2012)].

\bibitem{Shen:2010ky}
Lei-Lei Shen, Xiao-Lin Chen, Zhi-Gang Luo, Peng-Zhi Huang, Shi-Lin Zhu,
  Peng-Fei Yu, and Xiang Liu.
\newblock {The Molecular systems composed of the charmed mesons in the
  $H\bar{S}+\text{h.c.}$ doublet}.
\newblock {\em Eur. Phys. J. C}, 70:183--217, 2010.

\bibitem{PhysRevB.90.075128}
M.~Soriano and J.~J. Palacios.
\newblock Theory of projections with nonorthogonal basis sets: Partitioning
  techniques and effective hamiltonians.
\newblock {\em Phys. Rev. B}, 90(7):075128, 2014.

\bibitem{Cheng:2004ru}
Hai-Yang Cheng, Chun-Khiang Chua, and Amarjit Soni.
\newblock {Final state interactions in hadronic $B$ decays}.
\newblock {\em Phys. Rev. D}, 71:014030, 2005.

\bibitem{Faessler:2007gv}
Amand Faessler, Thomas Gutsche, Valery~E. Lyubovitskij, and Yong-Liang Ma.
\newblock {Strong and radiative decays of the $D_{s0}^*(2317)$ meson in the
  $DK$-molecule picture}.
\newblock {\em Phys. Rev. D}, 76:014005, 2007.

\bibitem{BaBar:2010tqo}
P.~del Amo~Sanchez et~al.
\newblock {Measurement of the $B \to \bar{D}^{(*)}D^{(*)}K$ branching
  fractions}.
\newblock {\em Phys. Rev. D}, 83:032004, 2011.

\bibitem{LHCb:2020qdy}
Roel Aaij et~al.
\newblock {Measurement of branching fraction ratios for $B^+\to D^{*+}D^-K^+$,
  $B^+\to D^{*-}D^+K^+$, and $B^0\to D^{*-}D^0K^+$ decays}.
\newblock {\em JHEP}, 12:139, 2020.

\bibitem{LHCb:2024vfz}
Roel Aaij et~al.
\newblock {Observation of New Charmonium or Charmoniumlike States in
  $B^+\to D^{*\pm}D^{\mp}K^+$ Decays}.
\newblock {\em Phys. Rev. Lett.}, 133(13):131902, 2024.

\bibitem{Pan:2025vyg}
Ya-Wen Pan, Ming-Zhu Liu, and Li-Sheng Geng.
\newblock {Probing the three-body force in hadronic systems with specific
  charge parity}.
\newblock arXiv:2512.01468 [nucl-th], 2025.

\end{thebibliography}
\end{document}